\documentstyle[12pt]{article}

\input{tcilatex}

\begin{document}

\title{Measurement, Filtering and Control\\
in Quantum Open Dynamical Systems}
\author{V P Belavkin \\
Mathematics Department\\
University of Nottingham\\
NG7 2RD UK\\
{\it email: vpb@maths.nott.ac.uk}}
\maketitle

\begin{abstract}
A Markovian model for a quantum automata, i.e. an open quantum dynamical
system with input and output channels and a feedback is described. A
multi-stage version of the theory of quantum measurement and statistical
decisions applied to the optimal control problem for quantum dynamical
discrete-time objects is developed. Quantum analogies of Stratonovich
non-stationary filtering and Bellman quantum dynamical programming for the
time being discrete are obtained.

The Gaussian case of quantum one-dimensional linear Markovian dynamical
system with a quantum linear transmission line is studied. The optimal
quantum multi-stage decision rule consisting of the classical linear optimal
control strategy and quantum optimal filtering procedure is found. The
latter contains the optimal quantum coherent measurement on the output of
the line and the recursive processing by Kalman--Busy filter.

All the results are illustrated by an example of the optimal control problem
for a quantum open oscillator at the input of a quantum wave transmission
line.
\end{abstract}


\section{Introduction}

High perspective of applying quantum coherent electromagnetic generators of
optical and infra-red frequency band for communication and control of
quantum dynamical objects stimulates an increase of the interest in
theoretical investigations of potential possibilities of information systems
containing quantum channels.

Due to fundamental limitations of quantum-mechanical measurement a specific
problem of optimal nondemolition measurement on the input and the output of
quantum channels arises in such investigations. Here we shall consider such
a problem for the channels with a feedback, corresponding to the optimal
control in quantum open systems. It is essential in quantum theory that
systems under the observation should be open, i.e. matched with channels, in
order not to demolish them, by letting out an information.

This paper gives the positive answers in a mathematically constructive way
to the following fundamental questions of quantum systems theory: Is it
possible at all to observe and control a quantum dynamical system in the
real time without not destroying it? If yes, what are the optimal strategies
of that observation and control? How the dynamics of a quantum system is to
be changed under the obtained information and its use as a feedback? What is
are the fundamental limitations of quantum observability and
controllability? Is there any possibility to obtain a time continuous limit
of such observation and control in a quantum system?

The non-dynamical problem of quantum measurement optimization formulated
primarily for detection and estimation in the static quantum communication
systems by K. Helstrom [1] was studied intensively by several authors [2--7]
within the framework of the single-stage (static) quantum-statistical
decision theory. The dynamic problem of quantum nondemolition measurement
for communication and control has been studied in details by author even in
continuous time [8,9] since the pioneering paper [10]. However the earlier
paper [11] on the solution of the discrete-time problem of optimal
measurement has never been published in full, in spite of the practical
importance of this case for the digital communication and control in quantum
channels with a feedback. The novelty of this paper was such, that only a
few people working in the newly open area of quantum stochastic processes
could appreciate it at that time, and it was too far yet from applications.
Recently, however, in view of the new possibilities of quantum computations,
the interest to quantum theory of communication and control has been
renewed, and the time development of the discrete models of quantum open
systems for communication and control became actual. Moreover, after the
development of time-continuous theory of quantum nondemolition measurement
and filtering within the quantum stochastic calculus approach [12], these
models can be considered as discrete time analogous and approximations of
this theory. The discrete time case is mathematically simpler, it doesn't
need the theory of quantum stochastic integration, and might be considered
on its own as a dynamical programming for quantum computations, or
multi-stage variant of optimal quantum-statistical decision theory. 

The quantum dynamical programming for multi-stage optimal measurement
problem can be considerably simplified due to assumption that not only the
processing of the measurement results but also the quantum measurement
itself may depend on all previous measurement results. It corresponds to the
assumption that we can choose a quantum measurement apparatus on the basis
of the previous measurement data separately at every instant in time. Though
in reality it is possible to imagine such a situation only for a finite
number of stages and a finite set of measurement results (time and measuring
scale being discrete), this extension of admissible measurement and decision
procedures is mathematically very convenient and from the physical point of
view is not contradictory. The choice of the measuring apparatus and of the
observed data processing according to all previous measurement results on
the whole defines the strategy in multi-stage quantum decision theory
described here. Within the framework of such an approach the problem of
quantum filtering of random signal sequences was reduced in [13] to the
well-studied problem of the static optimal quantum measurement on every
fixed stage with conditional a priori distribution depending on the previous
observed data.

Here we describe the multi-stage quantum statistical decision theory applied
to the problem of optimal control of a quantum Markovian discrete time
system with a matched quantum channel. This theory may be considered as a
quantum (operational) analogue of the stochastic control theory, based on
Stratonovich theory of conditional Markovian processes [14], and Bellman
dynamic programming [15].

The optimal filtering and the control strategy are found here in case of
one-dimensional quantum linear Markovian system with quantum Gaussian noises
and the mean-square loss function both in the discrete and continuous time.

In order to pose the problem of measurement and control correctly from the
physical point of view , let us consider the following motivating example.

\section{Controlled quantum open oscillator with quantum transmission line}

\setcounter{equation}{0} We are going to give a Markovian model of the
simplest quantum system with a communication channel: the quantum open
oscillator matched with a transmission line. It is an excellent mathematical
model of a single-mode antenna for quantum radiophysics and optical control
and communication.

Let $x$ be an operator of complex amplitude of a quantum oscillator with
Hamiltonian $\Omega x^{*}x$, which satisfies the canonical commutation
relations with $x^{*}$ being an adjoint operator

\begin{equation}
\left[ x,x^{*}\right] =xx^{*}-x^{*}x=\hbar {\bf 1}
\end{equation}
where ${\bf 1}$ is the unit operator, and $\hbar >0$ is the Planck constant.

Assume that in general case this oscillator is controlled by the complex
amplitude $u$ by means of a quantum-mechanical transmission line with wave
resistance $\gamma /2$, where the operator of the wave $y\left( t-\frac{s}{c}%
\right) $ travelling from the oscillator into the line is measured. In the
simplest case of ideal conjugation between the line and the measuring
apparatus, when there is no reflection of the wave travelling from the
oscillator, i.e.\thinspace in case of the matched line, $x\left( t\right) $
and $y\left( t\right) $ are described by the pair of linear equations [16] 
\begin{equation}
dx\left( t\right) /dt+\alpha x(t)=\gamma u(t)+v\left( t\right) ,\quad
\,x\left( 0\right) =x,
\end{equation}
\begin{equation}
y\left( t\right) =\bar{\alpha}x\left( t\right) -dx\left( t\right) /dt=\gamma
\left( x\left( t\right) -u\left( t\right) \right) -v\left( t\right) ,
\end{equation}
where, generally speaking, $\alpha $ is a complex number with fixed real
part, $\alpha +\bar{\alpha}=\gamma $, and with arbitrary imaginary part
depending on the choice of the representation, $v\left( t+\frac{s}{c}\right) 
$ the amplitude operator of the wave travelling out of line towards the
oscillator, this operator is responsible for the commutator preservation.
Under natural for super-high and optical frequencies assumption of
narrowness of the frequency band which we deal with the commutators for $%
v\left( t\right) $ in the representation of ``rotating waves'' have
delta-function form [17]:

\begin{equation}
\left[ v\left( t\right) ,v\left( t^{\prime }\right) \right] =0,\quad \left[
v\left( t\right) ,\,v\left( t^{\prime }\right) ^{*}\right] =\gamma \hbar 
{\bf 1}\delta \left( t-t^{\prime }\right) .
\end{equation}

Integrating equation (2.2) and taking into account that $v\left( t\right) $
does not depend on $x(t^{\prime })$ when $t>t^{\prime }$, it is easy to
verify that the commutator $\left[ x\left( t\right) ,x\left( t\right) ^{*}%
\right] $ is constant, moreover, $x\left( t\right) $ commutes both with $%
y\left( t^{\prime }\right) $ and $y\left( t^{\prime }\right) ^{*}$ when $%
t>t^{\prime }$, and the commutators for $y\left( t\right) ,y\left( t^{\prime
}\right) ,y\left( t^{\prime }\right) ^{*}$ coincide with (2.4). The latter
means that considering von Neumann reduction which appears as a result of
some quantum measurement of $y\left( t\right) $ at previous instants of time 
$t^{\prime }<t$ does not affect the future behaviour of $x\left( t^1\right)
,y\left( t^1\right) ,t^1>t$, so that equations (2.2), (2.3) remain
unchanged. This fact together with the Markovianity hypothesis of the
quantum process $x\left( t\right) $ which hold for quantum thermal
equilibrium states of the wave $v(t)$ in case of narrow band approximation
[17] simplifies to the large extent optimal measurement and control problems
for the simplest quantum dynamical system mentioned above.

Let us assume, that the initial state $x$ is Gaussian with the mathematical
expectation $\left\langle x\right\rangle =z$ and 
\[
\left\langle \left( x-z\right) \left( x-z\right) \right\rangle =0,\quad
\left\langle \left( x-z\right) ^{*}\left( x-z\right) \right\rangle =\hbar
\Sigma , 
\]
$v\left( t\right) $ is the quantum Gaussian white noise, which is described
by the following correlations 
\[
\left\langle v\left( t\right) \,v\left( t^{\prime }\right) \right\rangle
=0,\quad \left\langle v\left( t\right) ^{*}v\left( t^{\prime }\right)
\right\rangle =\hbar \sigma \delta \left( t-t^{\prime }\right) , 
\]
with $\sigma =\gamma \left( \exp \left( \hbar \Omega /kT\right) -1\right)
^{-1}$for the equilibrium state with the temperature $T$, where $k>0$ is the
Boltzmann constant.

As an example, let us try to choose the optimal measurement of the
controlled quantum oscillator (2.2) with transmission line (2.3), so that to
minimize its energy $\Omega \left\langle x^{\ast }\left( t\right) \,x\left(
t\right) \right\rangle $ at the final instant of time $t=\tau $ by means of
the control strategy the norm $\int_{0}^{\tau }\mid u\left( t\right) \mid
^{2}dt$ of which should not be too great. If we want also to force the
quantum amplitude $x(t)$ to follow the classical process $u(t)$, this
problem can be characterized by the quality criterion 
\begin{equation}
\Omega \left\langle x\left( \tau \right) ^{\ast }x\left( \tau \right)
\right\rangle +\int_{0}^{\tau }\left\langle \theta u\left( t\right) ^{\ast
}u\left( t\right) +\omega \left( x\left( t\right) -u\left( t\right) \right)
^{\ast }\left( x\left( t\right) -u\left( t\right) \right) \right\rangle dt.
\end{equation}
Here $\theta ,\omega \geq 0$ are parameters responsible for the measurement
quality: when $\theta =\Omega =0$ (2.5) corresponds to the problem of pure
filtration, when $\omega =0,\,\theta \neq 0$, it corresponds to the pure
control problem.

It will be shown below (see \S 5) that the optimal measurement minimising
criterion (2.5) is statistically equivalent to the measurement of the
stochastic process $z\left( t\right) =\widehat{x}\left( t\right) +x^{\circ
}\left( t\right) $ described by Kalman--Bucy filter: 
\begin{equation}
d\hat{x}\left( t\right) /dt+\alpha \hat{x}\left( t\right) =\gamma u\left(
t\right) +\kappa \left( t\right) \left( y\left( t\right) -\gamma \left( \hat{%
x}\left( t\right) -u\left( t\right) \right) \right) .
\end{equation}
Here $\hat{x}\left( 0\right) =z,\quad \kappa \left( t\right) =\left( \gamma
\Sigma \left( t\right) -\sigma \right) /\left( \mu +\sigma \right) ,\Sigma
\left( t\right) $ is the solution of the equation 
\[
d\Sigma \left( t\right) /dt=\left( \sigma -\gamma \Sigma \left( t\right)
\right) \left( \mu +\gamma \Sigma \left( t\right) \right) /\left( \mu
+\sigma \right) ,\quad \Sigma \left( 0\right) =\Sigma ,
\]
\begin{equation}
dx^{\circ }\left( t\right) /dt+\alpha x^{\circ }\left( t\right) =\kappa
\left( t\right) \left( v^{\circ }\left( t\right) -\gamma x^{\circ }\left(
t\right) \right) ,\quad x^{\circ }\left( 0\right) =0,
\end{equation}
where $v^{\circ }\left( t\right) $ is the amplitude operator with
commutators 
\[
\left[ v^{\circ }\left( t\right) ,\,v^{\circ }\left( t^{\prime }\right) %
\right] =0,\quad \left[ v^{\circ }\left( t\right) ,\,v^{\circ }\left(
t^{\prime }\right) ^{\ast }\right] =-\hbar \gamma \delta (t-t^{\prime })
\]
which change the quantum process $\hat{x}\left( t\right) $ into the
classical (commutative) diffusion complex process, and with correlations of
vacuum noise of the intensity $\mu =0$ if $\gamma \leq 0$ and $\mu =\gamma $
if $\gamma >0$: 
\begin{equation}
\left\langle v^{\circ }\left( t\right) v^{\circ }\left( t^{\prime }\right)
\right\rangle =0,\,\left\langle v^{\circ }\left( t\right) ^{\ast }v^{\circ
}\left( t^{\prime }\right) \right\rangle =\hbar \mu \delta \left(
t-t^{\prime }\right) .
\end{equation}
For instance, such measurement takes place by the heterodyning [7] where $%
v^{\circ }\left( t\right) $ stands for a standard wave. In this case the
optimal control strategy $u^{o}\left( t\right) $ coincides with the
classical one: $u^{o}\left( t\right) =-\lambda \left( t\right) \,z\left(
t\right) $, where $\lambda (t)=\left( \gamma \Omega \left( t\right) -\omega
\right) /\left( \theta +\omega \right) $, and $\Omega \left( t\right) $ is a
solution of the equation:

\begin{equation}
-d\Omega \left( t\right) /dt=\left( \omega -\gamma \Omega \left( t\right)
\right) \,\left( \theta +\gamma \Omega \left( t\right) \right) /\left(
\theta +\omega \right) ,\quad \Omega \left( \tau \right) =\Omega ,
\end{equation}
which together with (2.7) defines the minimum quantity of losses (2.5): 
\[
\hbar \left( \Omega \left( 0\right) \Sigma +\int_0^\tau \left( \Omega \left(
t\right) \sigma +\left( \gamma \Omega \left( t\right) -\omega \right)
^2\Sigma \left( t\right) /\left( \theta +\omega \right) \right) dt\right)
+\Omega \left( 0\right) \mid z\mid ^2. 
\]

By setting $\sigma =0,\,\omega =0$, we obtain in particular the solution of
the terminal control problem for an oscillator with thermal noise equal to
zero. But in this case unlike the classical one the optimal measurement
remains indirect and the equation (2.7) remains regular corresponding to the
white noise in the channel of intensity $\mid \gamma \mid \hbar $. Thus to
consider the quantum measurement postulates is statistically equivalent to
the adding of white noise into the channel of intensity $\mid \gamma \mid
\hbar $ what excludes the singular case of pure measurement of the amplitude 
$\hat{x}$.

It is interesting to note that in the case of thermal equilibrium when $%
\gamma >0,\,T>0$ and $\Sigma =\left( \exp \left\{ \hbar \Omega /kT\right\}
-1\right) ^{-1}$ the optimal amplification coefficient $\kappa \left(
t\right) $ equals to zero which means the possibility of optimal control of
the quantum oscillator without measurement. It also holds when $\omega
=\gamma \Omega $, the solution of equation (2.9) is stationary and optimal
feed-back coefficient $\lambda \left( t\right) $ equals to zero. But in the
contrary case $\gamma <0,\,T<0$ which corresponds to the active medium of
the oscillator (laser) the optimal coefficients $\kappa \left( t\right)
,\,\lambda \left( t\right) $ are strictly negative and non-zero even for the
stationary solution $\Sigma \left( t\right) =0,\,\Omega \left( t\right)
=\theta /\mid \gamma \mid $ of equations (2.7), (2.9).

\section{Quantum dynamical filtering}

\setcounter{equation}{0}

Now let us give a rigorous setting of the quantum dynamical observation
problem for the optimal control of a quantum-mechanical object when time is
discrete $t\in \left\{ t_k\right\} \,_{k=0,\,1,\ldots }$. Let ${\cal A}_k$
be von Neumann algebras on a Hilbert space ${\cal H}$, each is generated by
one or a few dynamical variables (operators) $x_k=\left( x_k^i\right) ^{i\in
I}$ in ${\cal H}$. One can consider a quantum-mechanical object in the
Heisenberg picture at the instant of time $t_{k+1}>t_k>0$ with $x_k=x\left(
t_k\right) $, such that all algebras ${\cal A}_k$ are equivalent to the
initial algebra ${\cal A}_0={\cal A}$, generated by the positions and
momentums $x=\left( q,p\right) $ at $t=0$. Let ${\cal B}_k,k=1,2,...$ be von
Neumann algebras of observables generated in ${\cal H}$ by output dynamical
variables $y_k=\left( y_k^j\right) ^{j\in J}$, by means of which this object
can be observed in a nondemolition way say, on the time intervals$%
\,(t_{k-1},t_k]$. As it has been shown above on the example of the matched
transmission line, the output observables $b_k\in {\cal B}_k$ in the matched
channels should commute with all present and future operators $a_{k^1}\in 
{\cal A}_{k^1},\,k^1\geq k$ of the dynamical system, but not necessarily
with the past ones $a_{k^{^{\prime }}}\in {\cal A}_{k^{^{\prime }}}$, $%
k^{^{\prime }}<k$. This commutativity condition together with the
commutativity $b_k^{\prime }b_{k^{\prime }}=b_{k^{\prime }}b_k^{\prime }$
for all $b_k^{\prime }\in {\cal B}_k$, $b_{k^{^{\prime }}}\in {\cal B}%
_{k^{^{\prime }}}$ $\forall k^{\prime }\neq k$ will be referred as the
nondemolition condition.

Let us denote ${\cal P}_k,\,{\cal R}_k$ the dual spaces to ${\cal A}_k,{\cal %
B}_k$ with respect to some standard pairings $<.,.>$, say the subspaces of
trace class operators $\pi _k\in {\cal A}_k,\,\rho _k\in {\cal B}_k$ which
are dual to the simple algebras of all bounded operators $a_k\in {\cal A}%
_k,\,b_k\in {\cal B}_k$ on the corresponding Hilbert spaces with respect to
the bilinear trace-forms 
\[
<\pi _k,\,a_k>={\rm tr}\,\left[ \pi _ka_k\right] ,\,<\rho _k,b_k>={\rm tr}%
\left[ \rho _kb_k\right] , 
\]
and denote ${\cal S}_k$ the corresponding subspace dual to the von Neumann
algebra ${\cal B}_k\,\vee {\cal A}_k$ generated by the commutating ${\cal B}%
_k$ and ${\cal A}_k$. We shall use the operational terminology, briefly
summarised in the Appendix. Thus we shall call the positive normalized
elements $\pi _k\in {\cal P}_k,\,\rho _k\in {\cal R}_k$ and $\sigma _k\in 
{\cal S}_k$, which are usually described by the statistical density
operators, the statistical states of the quantum object at the instants of
time $t_k$, the states of the channel on the interval $(t_{k-1},\,t_k]$, and
the joint state of the object and channel at the moment $t_k$ respectively,
or simply the states on ${\cal A}_k,\,{\cal B}_k$ and ${\cal B}_k\,\vee 
{\cal A}_k\subseteq {\cal B}_k\otimes {\cal A}_k$.

Now we adopt the hypothesis of Markovianity of the Heisenberg dynamics,
restricted to the described quantum object and output channel in ${\cal H}$,
with respect to a given state of the whole system $\omega $. Let all the
induced states $\sigma _{k}=\omega |\left( {\cal B}_{k}\vee {\cal A}%
_{k}\right) ,\,k=1,2\ldots $ and their restrictions $\rho _{k},\,\pi _{k}$
on ${\cal B}_{k},\,{\cal A}_{k}$ be defined by the initial state $\pi
_{0}=\pi $ on ${\cal A}_{0}={\cal A}$ and by a family $\left\{ M_{k}\right\}
\,_{k-1,2\ldots }$ of statistical morphisms $\pi _{k-1}\mapsto \sigma
_{k}=\pi _{k-1}M_{k}$. These transition maps ${\cal P}_{k-1}\rightarrow 
{\cal S}_{k}$ can be described as the pre-dual to positive normalized
superoperators $M_{k}:{\cal B}_{k}\otimes {\cal A}_{k}\rightarrow {\cal A}%
_{k-1}$ having for the simple algebras the form 
\[
M_{k\,}c_{k}={\rm tr}_{{\cal B}_{k}^{\circ }}\left[ \rho _{k}^{\circ }\,c_{k}%
\right] ,\quad \forall c_{k}\in {\cal B}_{k}\otimes {\cal A}_{k}.
\]
Here $\rho _{k}^{\circ },\,k=1,2,\ldots $ are states on some algebras ${\cal %
B}_{k}^{\circ }$, for which the simple algebras ${\cal B}_{k}\otimes {\cal A}%
_{k}$ are isomorphic to the von Neumann tensor products ${\cal A}%
_{k-1}\otimes {\cal B}_{k}^{\circ }$, and ${\rm tr}_{{\cal B}^{\circ }}$ is
the partial trace on ${\cal B}^{\circ }$ such that $M_{k}\left[
a_{k-1}\otimes b_{k}^{\circ }\right] =<\rho _{k}^{\circ },b_{k}^{\circ
}>a_{k-1}$ for all $a_{k-1}\in {\cal A}_{k-1},\,b_{k}^{\circ }\in {\cal B}%
_{k}^{\circ }$. This assumption corresponds to the requirement that the
channel should be matched with the object and implies the semigroup dynamics
[20] $\pi _{k-1}\mapsto \pi _{k}={\rm tr}_{{\cal B}_{k}}\left\{ \pi
_{k}\,M_{k}\right\} $ of the quantum-mechanical object with discrete time.
Furthermore, we shall suppose that every morphism $M_{k}$ may depend on the
results $\zeta ^{k}=\left\{ \zeta _{k^{\prime }}\right\} _{k^{\prime }<k}$
of previous measurement data $\zeta _{k^{\prime }}\in Z,k^{\prime }<k$, say
via dependence of some controlled parameters $u\in U$ of the sequence $%
\left\{ \zeta _{k^{\prime }}\right\} _{k^{\prime }<k}$ due to a feedback $%
\zeta ^{k}\mapsto u$. 

The nondemolition measurements during the time intervals $(t_{k-1},\,t_{k}]$
are described by positive operator-valued measures $b_{k}\,\left( d\zeta
\right) \in {\cal B}_{k},\,k=1,2,\ldots $ on the data space $Z\ni \zeta $
with a given Borel structure of the measurable subsets $dz\subseteq Z$ such
that $b_{k}\left( Z\right) ={\bf 1}$ is the identity operator of ${\cal B}%
_{k}$. We shall assume that every $Z$-measurement $b_{k}\left( d\zeta
\right) $ also may depend on all preceding measurement results $\zeta
_{1},\ldots ,\,\zeta _{k-1}$, and not only due to a dependance on $u\in U$
and the feedback, but directly, being adaptive in time. The functions $\zeta
^{k}\mapsto (M_{k}\left( \zeta ^{k}\right) ,\,b_{k}\left( \zeta ^{k},d\zeta
\right) )$ are supposed to be weakly measurable in the sense that for all $%
\pi _{k-1}\in {\cal P}_{k-1}$ and $a_{k}\in {\cal A}_{k}$ and all Borel
subsets $d\zeta \subseteq Z$ the complex functions 
\[
\zeta ^{k}\mapsto <\pi _{k-1}M_{k}\left( \zeta ^{k}\right) ,\,b_{k}\left(
\zeta ^{k},d\zeta \right) a_{k}>
\]
are Borel functions on $Z^{k}=\prod_{k^{\prime }<k}Z_{k^{\prime }},$ where $%
Z_{k}=Z,Z_{0}=U$. We shall call every sequence $\left\{ b_{k}\left( \zeta
^{k},d\zeta \right) \right\} _{k=1,2,\ldots }$ of such ``conditional'', or
adaptive measurements the measurement strategy.

Let us denote $B_k\left( \zeta ^k,d\zeta \right) $ the conditional
transition measures ${\cal P}_{k-1}\rightarrow {\cal P}_k$, that is the
operational-valued conditional measures on $Z$, defined as the predual to
superoperator values $B_k\left( \zeta ^k,d\zeta \right) :{\cal A}%
_k\rightarrow {\cal A}_{k-1}$ by the formula 
\begin{equation}
a_k\mapsto B_k\left( \zeta ^k,d\zeta \right) a_k=M_k\left( \zeta ^k\right) 
\left[ b_k\left( \zeta ^k,d\zeta \right) \otimes a_k\right] ,
\end{equation}
and denote $\pi _k\,\left( d\zeta ^{k+1}\right) $ the ${\cal P}_k$ - valued
measures on $Z^{k+1}$ obtained for $k=1,2,\ldots $ by the recurrency 
\begin{equation}
\pi _k\left( d\zeta ^k\times d\zeta \right) =\pi _{k-1}\left( d\zeta
^k\right) B_k\left( \zeta ^k,d\zeta \right)
\end{equation}
with the initial condition $\pi _0\left( d\zeta ^1\right) =\pi \delta \left(
u_0,d\zeta ^1\right) $ if $Z^1=U$.

\begin{lemma}
All the measures $\pi _k\left( d\zeta ^{k+1}\right) $ are positive in the
sense that 
\[
\int <\pi _k\left( d\zeta ^{k+1}\right) ,a_k\left( \zeta ^{k+1}\right)
>\,\geq 0 
\]
for all ${\cal A}_k$ - valued positive measurable functions $a_k\left( \zeta
^{k+1}\right) \geq 0$ and are normalized,

$<\pi _k\left( Z^{k+1}\right) ,{\bf 1}>=1$, where ${\bf 1}$ is the identity
operator of ${\cal A}_k$.
\end{lemma}

{\noindent \bf Proof } %
%
As the superoperator-valued measures $B_k\left( \zeta ^k,d\zeta \right) $
are positive and normalized in the sense that $\int B_k\left( \zeta
^k,d\zeta \right) \left[ a_k\left( \zeta ^k,\zeta \right) \right] \ge 0$ for
all $a_k\left( \zeta ^{k+1}\right) \ge 0$ and $B_k\left( \zeta ^k,Z\right) 
{\bf 1}={\bf 1}$, the lemma can be easily proved by induction, using the
positivity and normalization of $\pi _0$. Thus the measure $\pi _k\left(
d\zeta ^{k+1}\right) $, obtained by the recurrency (3.2), describes the
total statistical state on the algebra ${\cal A}_k$ and on the expanding
space $Z^{k+1}=Z^k\times Z$ $\vrule height.9exwidth.8exdepth-.1ex$

Let us define a posteriori state of the object at time $t_k$ as ${\cal P}_k$
--valued Radon-Nikodim derivative 
\begin{equation}
\pi _{k-1}\left( \zeta ^k\right) =\pi _{k-1}\left( d\zeta ^k\right) /<\pi
_{k-1}\left( d\zeta ^k\right) ,\,{\bf 1>}
\end{equation}
which exists in the weak sense due to absolute continuity of $\pi _{k-1}$
with respect to $<\pi _{k-1},{\bf 1}>$.

\begin{theorem}
The a posteriori states $\pi _k\left( \zeta ^{k+1}\right) ,\,k=1,2,\ldots $
can be obtained by the non-linear recurrency 
\[
\pi _k\left( \zeta ^k,\zeta \right) =\pi _{k-1}\left( \zeta ^k\right)
T_k\left( \zeta ^k,\zeta ,\pi _{k-1}\left( \zeta ^k\right) \right) ,\quad
\pi _0\left( \zeta ^1\right) =\pi , 
\]
where $T_k\left( \zeta ^{k+1},\pi _{k-1}\right) $ is the $\left( {\cal P}%
_{k-1}\rightarrow {\cal P}_k\right) $-valued Radon-Nikodim derivative 
\[
T_k\left( \zeta ^k,\zeta ,\pi _{k-1}\right) =B_k\left( \zeta ^k,d\zeta
\right) /<\pi _{k-1}B_k\left( \zeta ^k,d\zeta \right) ,{\bf 1}>. 
\]
\end{theorem}

{\noindent \bf Proof }%
%
The nonlinear transition operations $T_{k}$ are defined in the weak sense
almost everywhere by the Radon-Nikodim derivatives 
\[
<\pi _{k-1}T_{k}\left( \zeta ^{k+1},\pi _{k-1}\right) ,a_{k}>=<\pi
_{k-1}B_{k}\left( \zeta ^{k},d\zeta \right) ,a_{k}>/<\pi _{k-1}B_{k}\left(
\zeta ^{k},d\zeta \right) ,{\bf 1}>.
\]
The proof of the theorem follows immediately by induction due to the Bayes
formula 
\[
<\pi _{k}\left( d\zeta ^{k}\times d\zeta \right) ,{\bf 1>}\,/<\,\pi
_{k-1}\left( d\zeta ^{k}\right) ,\,{\bf 1}>=<\pi _{k-1}\left( \zeta
^{k}\right) B_{k}\left( \zeta ^{k},d\zeta \right) ,{\bf 1}>,
\]
from the definitions (3.2), (3.3) $\vrule height.9exwidth.8exdepth-.1ex$

Note that the equation (3.4), describing the conditional Markovian evolution
of a posteriori state of a quantum-mechanical object, can be regarded as a
quantum generalization of Stratonovich nonlinear filter equation with
discrete time. A semi-quantum case when a partially observed object is
described by a classical Markovian process $\left\{ x_k\right\}
_{k=0,1,\ldots }$ and the channel is non-classical, was considered in [12].

\section{Quantum dynamical programming}

\setcounter{equation}{0} Let us consider the problem of optimization of the
observation strategy $\left\{ b_k\left( \zeta ^k,\,d\zeta \right) \right\} $
on the fixed discrete time interval $\left[ 0,K\right] $. The optimal
strategy $\left\{ b_k^o\right\} _{k\in [0,K)}$ is defined as a strategy,
which minimizes the average cost 
\begin{equation}
\alpha =<\pi _K,\,a_K>+\sum_{k=1}^K\int <\pi _{k-1}\left( d\zeta ^k\right)
,c_{k-1}\left( \zeta ^k\right) >,
\end{equation}
given by a self-adjoint semi-bounded operator $a_K\in {\cal A}_K$ of final
losses, and by similar operator-valued functions $\zeta ^{k+1}\mapsto
c_k\left( \zeta ^{k+1}\right) \in {\cal A}_k$, $k=0,\ldots ,K-1$. (In the
case of unbounded $a_K$ and $c_k\left( \zeta ^{k+1}\right) $ only their
spectral measures should belong to ${\cal A}_K$ and ${\cal A}_k$.) Let us
remark that the cost (4.1) does not depend on the last measurement $%
b_K\left( \zeta ^K,\,d\zeta \right) $ which can be chosen arbitrarily, and $%
\pi _K=\pi _K\left( Z^{K+1}\right) $. As it follows from definitions (3.1),
(3.2) the $\sum_{k^{\prime }=1}^k$ in (4.1) for any $k=1,\ldots ,K$ is
independent of the measures $b_{k^1}\left( \zeta ^{k^1},d\zeta \right) $ for 
$k^1\geq k$. Hence in order to find the optimal $Z$--measurement $b_k\left(
\zeta ^k,d\zeta \right) $ from some $k<K$ it is enough to vary the future
average observation cost functional 
\begin{equation}
\alpha _k=<\pi _K,\,a_K>+\sum_{k^{\prime }=k+1}^K\int <\pi _{k^{\prime
}-1}\left( d\zeta ^{k^{\prime }}\right) ,\,c_{k^{\prime }-1}\left( \zeta
^{k^{\prime }}\right) >.
\end{equation}

\begin{lemma}
The explicit dependence of $\alpha _k$ on $b_k\left( \zeta ^k,\,d\zeta
\right) $ is affine 
\begin{equation}
\alpha _k=\int_{Z^k}\int_Z\,<\rho _k\,\left( d\zeta ^k,\,\zeta \right)
,\,b_k\left( \zeta ^k,\,d\zeta \right) >,
\end{equation}
where $\rho _k\left( d\zeta ^k,\zeta \right) =\pi _{k-1}\left( d\zeta
^k\right) A_k\left( \zeta ^k,\zeta \right) $. Here $A_k\left( \zeta
^{k+1}\right) $ is a $\left( {\cal P}_{k-1}\rightarrow {\cal R}_k\right) $
--valued function on $\zeta ^{k+1}$ which is defined as predual to the
superoperators 
\begin{equation}
b_k\mapsto A_k\left( \zeta ^{k+1}\right) b_k=M_k\left( \zeta ^k\right) \left[
b_k\otimes \,a_k\left( \zeta ^{k+1}\right) \right] ,\quad \forall b_k\in 
{\cal B}_k,
\end{equation}
where $a_k\left( {\cal \zeta }^{k+1}\right) $ is an operator-valued function
on $Z^k$ satisfying the linear inverse-time recurrency 
\begin{equation}
a_{k-1}\left( \zeta ^k\right) =\int B_k\left( \zeta ^k,d\zeta \right)
\,a_k\left( \zeta ^k,\zeta \right) +c_{k-1}\left( \zeta ^k\right) ,
\end{equation}
$k=1,...,K$ with the boundary condition $\alpha _K\left( \zeta ^{K+1}\right)
=a_K$.
\end{lemma}

{\noindent \bf Proof } %
%
First let us prove that the future losses (4.2) can be represented as 
\[
\alpha _k=\int_{Z^k}\int_Z<\pi _k\left( d\zeta ^k\times d\zeta \right)
,a_k\left( \zeta ^k,\zeta \right) >, 
\]
where $a_k\left( \zeta ^{k+1}\right) \in {\cal A}_k$ is the solution to the
equation (4.5). It is obviously valid for $k=K$, and if it is true for a $%
k<K $, then substituting (3.2) into this representation of $\alpha _k$, we
obtain 
\begin{eqnarray*}
&&\int <\pi _k\left( d\zeta ^{k+1}\right) ,a_k\left( \zeta ^{k+1}\right)
>+\int <\pi _{k-1}\left( d\zeta ^k\right) ,c_{k-1}\left( \zeta ^k\right) > \\
&=&\int_{Z^k}<\pi _{k-1}\left( d\zeta ^k\right) ,\int_ZB_k\left( \zeta
^k,d\zeta \right) a_k\left( \zeta ^k,\zeta \right) >+c_{k-1}\left( \zeta
^k\right) .
\end{eqnarray*}
So this is also valid for $\alpha _{k-1}$ with $a_{k-1}$ given in (4.5), and
by using the inverse-time induction, it is valid for any $k\in [0,K)$. Now
we can obtain (4.3) by 
\[
<\pi _k\left( d\zeta ^k\times d\zeta \right) ,a_k\left( \zeta ^k,\zeta
\right) >=<\pi _{k-1}\left( d\zeta ^k\right) B_k\left( \zeta ^k,d\zeta
\right) ,a_k\left( \zeta ^k,\zeta \right) > 
\]
\[
=<\pi _{k-1}\left( d\zeta ^k\right) M_k\left( \zeta ^k\right) ,b_k\left(
\zeta ^k,d\zeta \right) \otimes a_k\left( \zeta ^k,\zeta \right) > 
\]

\[
=<\pi _{k-1}\left( d\zeta ^k\right) A_k\left( \zeta ^k\right) ,b_k\left(
\zeta ^k,d\zeta \right) >=<\rho _k\left( d\zeta ^k,\zeta \right) ,b_k\left(
\zeta ^k,d\zeta \right) >, 
\]
where we used the definitions (3.1) and (4.4) for the operations $B_k$ and $%
A_k$ $\vrule height.9exwidth.8exdepth-.1ex$

\begin{theorem}
If the strategy $\left\{ b_k^o\left( \zeta ^k,d\zeta \right) \right\} _{k\in
[1,K)}$ is optimal for the cost functional (4.1), it satisfies the following
system of equations 
\begin{equation}
\left( \rho _k\left( d\zeta ^k,\zeta \right) -\lambda _k\left( d\zeta
^k\right) \right) b_k^o\left( \zeta ^k,d\zeta \right) =0,
\end{equation}
$k\in [1,K)$, where 
\[
\lambda _k\left( d\zeta ^k\right) =\int_Z\rho _k\left( d\zeta ^k,\zeta
\right) b_k^o\left( \zeta ^k,d\zeta \right) . 
\]
These equations together with the system of inequalities 
\begin{equation}
\rho _k\left( d\zeta ^k,\zeta \right) \ge \lambda _k\left( d\zeta ^k\right)
,\quad k\in [1,K)
\end{equation}
give the necessary and sufficient conditions of the optimality for quantum
measurement strategy $b_k^o$, $k=1,...K-1$ corresponding to the minimal
values 
\begin{equation}
\alpha _k^o=\int <\lambda _k\left( d\zeta ^k\right) ,\,{\bf 1}>
\end{equation}
of the future average costs (4.2).
\end{theorem}

{\noindent \bf Proof } %
%
As the variables $b_k\left( \zeta ^k,d\zeta \right) ,\,k=1,2,\ldots ,K$ of
the functional (4.2) are independent, the optimal measure $b_k^o\left( \zeta
^k,d\zeta \right) $ minimizes the affine functional separately for every
fixed family $\left\{ b_{k^1}\left( \zeta ^{k^1},d\zeta \right) \right\}
_{k^1>k}$. The necessary and sufficient conditions (4.6), (4.7) of
optimality for $b_k^o\left( \zeta ^k,d\zeta \right) $, minimizing the affine
functional (4.3) with a fixed $k$, follow immediately by the linear
programming method, as it was noted in the single-stage theory of optimal
quantum measurements [2, 4--7] $\vrule height.9exwidth.8exdepth-.1ex$

Note that the minimal value $\alpha ^o$ of the total average cost (4.1) is
given by the solution $a^o=a_0^o$ of the recurrency (4.5) with $B_k=B_k^o$
at $k=0$ as $\alpha ^o=<\pi ,a^o>$.

Let us note that with the help of the a posteriori states $\pi _k\left(
\zeta ^k\right) $, one can write conditions (4.6), (4.7) in the following
form 
\begin{equation}
\left( \rho _k\left( \zeta ^k,\zeta \right) -\lambda _k\left( \zeta
^k\right) \right) b_k^o\left( \zeta ^k,d\zeta \right) =0,
\end{equation}
\begin{equation}
\rho _k\left( \zeta ^{k+1}\right) \ge \lambda _k\left( \zeta ^k\right)
,\quad k\in [1,K),
\end{equation}
where $\rho _k\left( \zeta ^{k+1}\right) =\pi _{k-1}\left( \zeta ^k\right)
A_k\left( \zeta ^{k+1}\right) $. According to the Bellman dynamical
programming method [15] the verification of the optimality condition
formulated above can be carried out sequentially in inverse time $%
k=K-1,\ldots ,1$ applying the recurrence (4.5) for the superoperator $%
A_k\left( \zeta ^{k+1}\right) $ after solving the filtering recurrent
equation (3.4).

The optimal control of Markovian partially observed quantum-mechanical
object can be reduced to the optimal measurement problem investigated above
as follows. Let $M_k\left( u_{k-1}\right) :{\cal P}_{k-1}\rightarrow {\cal R}%
_k\otimes {\cal P}_k$ be the quantum statistical morphisms (transitions)
controlled by some parameters $u_k\in U,\,k=0,\ldots ,K-1$. A control
strategy $\left\{ \gamma _k\right\} _{k<K}$ is given by a choice of the
feedback, defined by a measurable dependence $\gamma _k$ of each $u_k$ on
all measurement data $\eta _{k^{\prime }}\in Y,\,k^{\prime }\le k$, and also
on the preceding controls $u_{k^{\prime }},\,k^{\prime }<k$. The optimal
control for a fixed measurement strategy is supposed to minimize the average
cost defined by a final operator $a_K$ and operator-valued cost functions $%
c_k\left( u_k\right) ,\,k=0,\ldots ,K-1$. Denoting $\zeta ^1=u_0$,$\,\zeta
^k=\left( u_0,\eta _1,\,u_1,\ldots ,\,\eta _{k-1},u_{k-1}\right) $, $\zeta
=\left( \eta ,u\right) ,$ the average cost functional even with random
control strategies can be represented in the form (4.1), given by the
quantum measurement strategy $\left\{ b_k\left( \zeta ^k,d\zeta \right)
\right\} $ on $Z=Y\times U$ of the form 
\begin{equation}
b_k^o\left( \zeta ^k,\,d\eta \times du\right) =b_k^o\left( \zeta ^k,\,d\eta
\right) \delta \left( \gamma _k^o\left( \zeta ^k,\,\eta \right) ,\,du\right)
\end{equation}
and $c_0\left( \zeta ^1\right) =c_0\left( u_0\right) $, $c_k\left( \zeta
^{k+1}\right) =c_k\left( u_k\right) $. The quantum optimal control problem
can be formulated then as one of searching for the optimal $Y\times U$
--measurements $b_k^o\left( \zeta ^k,d\zeta \right) ,\,k\in \left(
1,k\right) ,$ and an optimal initial control $u^o$ corresponding to the
minimal value 
\[
\alpha ^o=\inf_u<\lambda _1\left( u\right) ,{\bf 1}>+<\pi _0,c_0\left(
u\right) > 
\]
of average cost (4.1). In general, the optimal measurement strategy may not
be in the product form (4.11), but if there exists a non-randomized strategy 
$u_k^o=\gamma _k^o\left( \zeta ^k,\,\eta \right) ,\,k\in [1,\,K)$ for some $%
Y $-measurements $b_k^o\left( \zeta ^k,\,d\eta \right) $ for which the $%
Y\times U$ - measurements are optimal, where $\delta \left( \cdot ,\cdot
\right) $ is the Dirac $\delta $- measure, then the data spaces $Y$ may be
called the sufficient spaces. The optimal measurements $b_k^o\left( \zeta
^k,\,d\eta \right) $ on sufficient data spaces $Y$ satisfy obviously the
equations 
\[
\left( \rho _k\left( \zeta ^k,\,\eta ,\,\gamma _k^o\left( \zeta ^k,\,\eta
\right) \right) -\lambda _k\left( \zeta ^k\right) \right) \,b_k^o\left(
\zeta ^k,\,d\eta \right) =0,\quad k\in [1,\,K), 
\]
where 
\[
\lambda _k\left( \zeta ^k\right) =\int \rho _k\left( \zeta ^k,\,\eta
,\,\gamma _k^o\left( \zeta ^k,\,\eta \right) \right) b_k^o\left( \zeta
^k,\,d\eta \right) , 
\]
which together with the inequalities (4.10) are necessary and sufficient for
the non-randomized control strategy $\left\{ \gamma _k^o\right\} $.

\section{Quantum filtering in Boson linear Markovian system in a Gaussian
state}

\setcounter{equation}{0} We examine a Markovian one-dimensional quantum
dynamical system, described at discrete instants $t_k=k\Delta $ by the
algebras ${\cal A}_k$ and ${\cal B}_k$, which are$\,$ generated by the
non-selfadjoint operators $x_k\neq x_k^{*}$ and $y_k\neq y_k^{*}$
respectively, satisfying the canonical commutation relations. Let us suppose
that they act in the same Hilbert space ${\cal H}$, where they satisfy the
linear quantum stochastic equations 
\begin{equation}
x_k=\phi x_{k-1}+\beta u_{k-1}+v_k
\end{equation}
\begin{equation}
y_k=\gamma x_{k-1}+\delta u_{k-1}+w_k.
\end{equation}
Here $\phi ,\beta ,\gamma ,\delta $ are some, in general complex parameters,
the controls $u_k$ can also accept complex values, $x_0=x$ is the initial
operator in ${\cal H}$, generating the algebra ${\cal A}$, and $v_k,w_k $
are some operators in ${\cal H},$ generating the algebras ${\cal B}_k^{\circ
}$. To obtain the Markov dynamics, we need to assume the independence of $x$
and all the pairs $\left( v_k,w_k\right) $ such that the algebras ${\cal A}$
and ${\cal B}_k^{\circ }$, corresponding to different instants of time $t_k$%
, commutate, and the joint state $\omega $ is the product of the states on $%
{\cal A}$ and all ${\cal B}_k^{\circ },k=1,2\ldots $. We shall define the
canonical commutation relations for the generating operators $x,v_k,w_k$
with their adjoints as following: 
\[
\left[ x,x^{*}\right] =\hbar {\bf 1}\qquad \left[ v_k,v_k^{*}\right] =\left(
1-\mid \phi \mid ^2\right) \hbar {\bf 1}, 
\]
\begin{equation}
\left[ w_k,w_k^{*}\right] =\left( \varepsilon -\mid \gamma \mid ^2\right)
\hbar {\bf 1},\quad \left[ w_k,v_k^{*}\right] =-\,\bar{\phi}\gamma \hbar 
{\bf 1},
\end{equation}
where $\hbar >0$ and ${\bf 1}$ is the identity in ${\cal H}$ ( other,
unwritten commutators, including all those corresponding to different
instants of time to be equal to zero.) Here the choice of the commutator $%
\left[ w_k,v_k^{*}\right] $, responsible for the commutativity $\left[
y_k,x_k^{*}\right] =0$ is essential, the other nonzero commutators are
chosen so that the commutators 
\[
\left[ x_k,x_k^{*}\right] =\hbar {\bf 1},\quad \left[ y_k,y_k^{*}\right]
=\varepsilon \hbar {\bf 1} 
\]
should be constant. The described system we shall call the discrete linear
Markovian quantum open oscillator.

Let us describe the states $\pi _k\in {\cal P}_k$ by the Glauber [21]
distributions $p_k\left( \xi \right) $, $\xi \in {\bf C}$, normalized on the
complex plane ${\bf C}$ with respect to the Lebesgue measure $d\xi =d{\rm Re}%
\xi d{\rm Im}\xi /\pi \hbar $. In the representation described in the
Appendix, the Markovian morphisms ${\cal P}_{k-1}\rightarrow {\cal P}_k$,
corresponding to the linear equations (5.1), (5.2), transform the
distributions $p_{k-1}\left( \xi \right) $ into the two-dimensional
distributions 
\begin{equation}
g_k\left( \xi ,\eta \right) =\int q_k\left( \xi -\phi \xi ^1-\beta u,\eta
-\gamma \xi ^1-\delta u\right) p_{k-1}\left( \xi ^1\right) d\xi ^1,
\end{equation}
where $q_k\left( \xi ,\eta \right) $ are some other (not necessarily
Glauber) distributions on ${\bf C}^2$, which describe the independent states 
$\rho _k^{\circ }$ on algebras ${\cal B}_k^{\circ }$.

When $\varepsilon =0$, the operators $y_k,y_k^{*}$ are simultaneously
measurable, and the a posteriori states on ${\cal A}_k$ under the fixed
spectral values $y_{k^{\prime }}=\eta _{k^{\prime }}$ and $u_{k^{\prime }}$, 
$k^{\prime }<k$ are defined recurrently by the a posteriori Glauber
distribution $p_k\left( \xi \mid \zeta ^{k-1}\right) $ according the Bayes
formula 
\[
p_k\left( \xi \mid \zeta ^{k-1}\right) =g_k\left( \xi ,\eta _k\mid \zeta
^k\right) /r_k\left( \eta _k\mid \zeta ^k\right) . 
\]
Here $g_k\left( \xi ,\eta \mid \zeta ^k\right) $are the distributions
obtained by substitution of $p_{k-1}\left( \xi \mid \zeta ^k\right) $ into
(5.4) instead of $p_{k-1}\left( \xi \right) $, and 
\[
r_k\left( \eta \mid \zeta ^k\right) =\int g_k\left( \xi ,\eta \mid \zeta
^k\right) d\xi 
\]
are the probability distributions, describing the complex values $\eta _k$,
which arise as the results of the direct measurements of $y_k$ under the
fixed $\zeta ^k=\left( u_0,\eta _1,u_1,\ldots ,\eta _{k-1},u_{k-1}\right) $.

When $\varepsilon \neq 0$, only indirect measurement of $y_k$ are possible
which are described, for instance, by the ${\cal B}_k$ --valued measures 
\begin{equation}
b_k\left( d\eta \right) =\#m_k\left( \eta -y_k\right) \#d\eta ,
\end{equation}
represented by some distributions $m_k\left( \eta \right) $ on ${\bf C}$ as
it is described in the Appendix (A.3). In this case in order to calculate a
posteriori Glauber distribution one should change $q_k\left( \xi ,\eta
\right) $ in formula (5.4) for the distribution 
\begin{equation}
q_k^1\left( \xi ,\eta \right) =\int m_k\left( \eta -\eta ^1\right) q_k\left(
\xi ,\eta ^1\right) d\eta ^1.
\end{equation}

\begin{theorem}
Let the initial state $\pi $ of the quantum oscillator be described by the
Glauber distribution of Gaussian type 
\begin{equation}
p\left( \xi \right) =\frac 1\Sigma \exp \left\{ -\mid \xi -z\mid ^2/\hbar
\Sigma \right\} ,
\end{equation}
the distributions $q_k\left( \xi ,\eta \right) $, describing the transitions
(5.4), be also Gaussian: 
\begin{equation}
q_k\left( \xi ,\eta \right) =\frac{\exp \left\{ -\left( \nu \mid \xi \mid
^2+2{\rm Re}\upsilon \xi \bar{\eta}+\sigma \mid \eta \mid ^2\right) /\hbar
\left( \sigma \nu -\mid \upsilon \mid ^2\right) \right\} }{\sigma \nu -\mid
\upsilon \mid ^2},
\end{equation}
and the measures $b_k$ are described as in (5.5), by the Gaussian
distributions 
\begin{equation}
m_k\left( \eta \right) =\frac 1\mu \exp \left\{ -\mid \eta \mid ^2/\hbar \mu
\right\} .
\end{equation}
Then a posteriori states (3.3) at each instant $k=1,2,\ldots ,$ are given by
the conditional Glauber distributions of Gaussian form 
\begin{equation}
p_k\left( \xi \mid \zeta ^{k+1}\right) =\frac 1{\Sigma _k}\exp \left\{ -\mid
\xi -z_k\mid ^2/\hbar \Sigma _k\right\} ,
\end{equation}
where $z_k,\Sigma _k$ are defined by the recurrent equations of the complex
Kalman filter: 
\begin{equation}
z_k=\phi z_{k-1}+\beta u_{k-1}+\kappa _k\left( \eta _k-\gamma z_{k-1}-\delta
u_{k-1}\right) ,\quad z_0=z,
\end{equation}
\begin{equation}
\Sigma _k=\mid \phi \mid ^2\Sigma _{k-1}+\sigma -\mid \kappa _k\mid ^2\Psi
_k,\quad \Sigma _0=\Sigma ,
\end{equation}
where 
\[
\kappa _k=\left( \phi \bar{\gamma}\Sigma _{k-1}-\upsilon \right) /\Psi
_k\quad \Psi _k=\mid \gamma \mid ^2\Sigma _{k-1}+\nu ^1,\quad \nu ^1=\nu
+\mu . 
\]
\end{theorem}

{\noindent \bf Proof } %
%
Due to the chosen representation, the proof is similar to the derivation of
the classical one-dimensional Kalman filter for the complex Gaussian process 
$x_k$ given by (5.1) and $y_k^1=y_k+w_k^{\circ }$, where $w_k^{\circ }$ are
independent Gaussian variables with zero mean values and the covariances $%
\mu \geq \varepsilon $. (For this proof see, for instance, [22].) One should
only take into account that distributions (5.6) are also Gaussian of the
type (5.8) with the parameter $\nu ^1=\nu +\mu $ instead of $\nu $.
Substituting $q\left( \xi ,\eta \right) $ in (5.4) by $q^1\left( \xi ,\eta
\right) $ and $p_{k-1}\left( \xi \right) $ by the conditional distribution $%
p_{k-1}\left( \xi \mid \zeta ^k\right) $ of type (5.10), we obtain 
\[
g_k^1\left( \xi ,\eta \mid \zeta ^k\right) =p_k\left( \xi \mid \zeta
^{k-1}\right) r_k^1\left( \eta \mid \zeta ^k\right) , 
\]
where $p_k\left( \xi \mid \zeta ^{k-1}\right) $ is the Gaussian distribution
(5.10) with the parameters (5.11), (5.12), and 
\begin{equation}
r_k^1\left( \eta \mid \zeta ^k\right) =\frac 1{\Psi _k}\exp \left\{ -\mid
\eta -\gamma z_{k-1}\mid ^2/\hbar \Psi _k\right\} .
\end{equation}
Thus the quantum Gaussian filtering is controlled by the classical Kalman
filter for the complex amplitude in the Glauber representation $\vrule %
height.9exwidth.8exdepth-.1ex$

Note, that in distinction from the classical case, the covariance matrix of
distributions (5.8), (5.9) should not only be non--negative definite but
should also satisfy the Heisenberg uncertainty principle 
\begin{equation}
\left( 
\begin{array}{cc}
\sigma & -\upsilon \\ 
-\bar{\upsilon} & \nu
\end{array}
\right) \ge \left( 
\begin{array}{cc}
\mid \phi \mid ^2-1 & \phi \bar{\gamma} \\ 
\gamma \bar{\phi} & \mid \gamma \mid ^2-\varepsilon
\end{array}
\right) ,\quad \mu \ge \varepsilon ,
\end{equation}
as it follows from inequality (A.5). In particular it excludes the case $\mu
=0$ of the direct observation of $y_k$ when $\varepsilon >0$.

As shown in the next paragraph, a posteriori mathematical expectations $z_k$
with $\mu =\max \left( 0,\varepsilon \right) $ appear to be the optimal
estimates $u_k^o=z_k$ of the operators $x_k$ with respect to the square
quality criterion $c_k\left( u_k\right) =:\mid x_k-u_k\mid ^2:$ with the
minimal mean square error $\hbar \Sigma _k$. In the commutative case $\left[
x_k,x_k^{*}\right] =0$ this optimality was proved in [11].

Note, that instead of calculating $z_k$ by means of the recurrent formula
(5.11) using the results $\left( \eta _1,...,\eta _k\right) $ of the
indirect measurement (5.5) one may regard $z_k$ itself as a results of the
measurement described by the ${\cal B}_k$--valued measure: 
\begin{equation}
b_k\left( \zeta ^k,dz\right) =\#n_k\left( z-\hat{x}_k\right) \#dz,
\end{equation}
where 
\[
n_k\left( z\right) =\frac 1{\mid \kappa _k\mid ^2}m_k\left( z/\mid \kappa
_k\mid \right) , 
\]
and 
\begin{equation}
\hat{x}_k=\phi z_{k-1}+\beta u_{k-1}+\kappa _k\left( y_k-\gamma
z_{k-1}\right)
\end{equation}
is an operator, depending on the values $z_{k-1},u_k,$ and independent of
the preceding measurement and control results.

It is interesting to consider the time continuous limit, when the quantum
oscillator (5.1), (5.2) is described by the quantum stochastic differential
equations 
\begin{equation}
dx\left( t\right) +\alpha x\left( t\right) dt=\beta u\left( t\right)
dt+v\left( dt\right) ,
\end{equation}
\begin{equation}
y\left( dt\right) =\gamma x\left( t\right) dt+\delta u\left( t\right)
dt+w\left( dt\right) ,
\end{equation}
i.e.\thinspace by equations (5.1), (5.2) with $x\left( t_k\right) =x_k,\quad
y\left( \Delta t_k\right) =y_k,\quad \phi \simeq 1-\alpha \Delta ,\quad
\beta \simeq \beta \Delta ,\quad \gamma \simeq \gamma \Delta ,\quad
\varepsilon \simeq \varepsilon \Delta ,$ where $\left( \Delta t_k\right)
=t_k-t_{k-1}=\Delta $ tends to zero. In addition to that the commutation
relations (5.3) change in the following way 
\[
\left[ x,x^{*}\right] =\hbar {\bf 1},\quad \left[ v\left( dt\right) ,v\left(
dt\right) ^{*}\right] =\left( \alpha +\bar{\alpha}\right) \hbar dt{\bf 1}, 
\]
\[
\left[ w\left( dt\right) ,w\left( dt\right) ^{*}\right] =\varepsilon \hbar dt%
{\bf 1},\quad \left[ w\left( dt\right) ,\nu \left( dt\right) ^{*}\right]
=-\gamma \hbar dt{\bf 1}, 
\]
and the other commutators including those corresponding to the different
instants of time are equal to zero. By passing to the limit as $\Delta
\longrightarrow 0$ when $\sigma \simeq \sigma \Delta ,\quad \upsilon \simeq
\upsilon \cdot \Delta ,\quad \nu \simeq \nu \cdot \Delta $, it is easy to
obtain under the assumptions of the Theorem 3 that a posteriori state $\pi
\left( t,\zeta ^t\right) $ is described by the Glauber distribution $p\left(
t,\xi \mid \zeta ^t\right) $ of Gaussian type (5.10) with the parameters $%
z\left( t\right) ,\Sigma \left( t\right) $ which correspond to the
Kalman--Busy filter 
\begin{equation}
dz\left( t\right) +\alpha z\left( t\right) dt=\kappa \left( t\right) \left(
\eta \left( dt\right) -\left( \gamma z\left( t\right) -\delta u\left(
t\right) \right) dt\right) .
\end{equation}
Here $\kappa \left( t\right) =\left( \bar{\gamma}\Sigma \left( t\right)
-\upsilon \right) /\nu ^1,\quad \nu ^1=v+\mu ,\quad z\left( 0\right)
=z,\quad \Sigma \left( 0\right) =\Sigma ,\quad $%
\[
d\Sigma \left( t\right) /dt+\left( \alpha +\bar{\alpha}\right) \Sigma \left(
t\right) =\sigma -\mid \chi \left( t\right) \mid ^2\nu ^1, 
\]
and $\eta \left( dt\right) $ are the results of the corresponding indirect
measurement of $y\left( dt\right) $ which are realized by the measurement of
the sum $y\left( dt\right) +w^{\circ }\left( dt\right) $, where $w^{\circ
}\left( dt\right) $ is an independent quantum white noise, defined by the
coefficients $\varepsilon ,\mu :$%
\[
\left[ w^{\circ }\left( dt\right) ,w^{\circ }\left( dt\right) ^{*}\right]
=-\varepsilon \hbar dt{\bf 1},\quad \left\langle w^{\circ }\left( dt\right)
^{*}w^{\circ }\left( dt\right) \right\rangle =\mu \hbar dt. 
\]

As shown at the end of the next paragraph, such ``continuous'' measurement
appears to be also optimal in the Gaussian case when $\mu =\max \left(
0,\varepsilon \right) $.

\section{Optimal measurement and control for quantum open linear system}

\setcounter{equation}{0} In the following theorem it is not required that
the distributions $p_0,q_k$ and $m_k$ should be Gaussian and it is assumed
only that they should have the zero mathematical expectations, and the
covariations should coinside with the covariances $\Sigma ,\delta ,\upsilon
,\nu ,\mu $ of the distributions (5.7) -- (5.9) respectively, not necessary
being of the form (5.11).

\begin{theorem}
Let the operator of final losses be quadratic: $a_K=\Omega x_K^{*}x_K$,
where $\Omega \ge 0$, and 
\begin{equation}
c_k\left( u_k\right) =\omega x_k^{*}x_k-\vartheta \bar{u}_kx_k-\bar{\vartheta%
}u_kx_k^{*}+\vartheta ^1\mid u_k\mid ^2,\quad \omega \ge 0,\quad \vartheta
^1>0
\end{equation}
be quadratic loss operators for all $k\in [0,K)$. Suppose $u_k=-\lambda
_kz_k,k\in [0,K)$ is a linear control strategy, where $z_k$ are the linear
estimates (5.11) based on the results $\eta _k$ of the indirect measurement
(5.5), and 
\begin{equation}
\lambda _k=\left( \phi \bar{\beta}\Omega _{k+1}-\vartheta \right) /\Upsilon
_k,
\end{equation}
with $\Upsilon _k=\mid \beta \mid ^2\Omega _{k+1}+\vartheta ^1$ and $\Omega
_k$ satisfying the following equation 
\begin{equation}
\Omega _k=\mid \phi \mid ^2\Omega _{k+1}+\omega -\mid \lambda _k\mid
^2\Upsilon _k,\quad \Omega _K=\Omega .
\end{equation}
Then the operators of future losses (4.5) are also quadratic: 
\[
a_k\left( \zeta ^{k+1}\right) =d_k{\bf 1}+\Upsilon _k\mid u_k+\lambda
_kz_k\mid ^2+\Omega _kx_k^{*}x_k 
\]
\begin{equation}
+\Gamma _k\left( z_k-x_k\right) ^{*}\left( z_k-x_k\right) -2{\rm Re}\Lambda
_k\left( u_k+\lambda _kz_k\right) ^{*}\left( z_k-x_k\right) ,
\end{equation}
where 
\[
d_k=\hbar \sum_{i=k+1}^K\left( \Omega _i\sigma +\Gamma _i\left( \sigma +2%
{\rm Re}x_i\bar{\upsilon}+\nu ^1\mid \kappa _i\mid ^2\right) \right) , 
\]
\begin{equation}
\Gamma _k=\mid \lambda _k\mid ^2\Upsilon _k+\mid \phi -\kappa _{k+1}\gamma
\mid ^2\Gamma _{k+1},\quad \Gamma _k=0,
\end{equation}
and 
\[
\Lambda _k=\phi \bar{\beta}\Omega _{k+1}-\vartheta . 
\]
\end{theorem}

{\noindent \bf Proof } %
%
In the representation 
\[
a_K=:\alpha _K\left( x_K\right) :,\quad c_k\left( u_k\right) =:\sigma \left(
x_k,u_k\right) :,\quad a_k\left( \zeta ^{k+1}\right) =:\alpha _{k+1}\left(
x_k,\zeta ^{k+1}\right) : 
\]
the recurrent equation (4.6) has the form 
\begin{equation}
\alpha _k\left( \xi _k,\zeta ^{k+1}\right) =\int \alpha _{k+1}\left( \xi
,\zeta ^{k+1},\eta ,u\right) q^1\left( \xi -\phi \xi _k-\beta u_k,\eta
-\gamma \xi _k-\delta u_k\right) d\xi d\eta +\sigma \left( \xi _k,u_k\right)
\end{equation}
where 
\[
u=-\lambda _{k+1}z,\quad z=\phi z_k+\beta u_k+\kappa _{k+1}\left( \eta
-\gamma z_k\right) . 
\]
Let us assume that the function $\alpha _k\left( \xi \right) $ has the
quadratic form (6.4); in particular, it has this form at $k=K$, namely $%
\alpha _K\left( \xi \right) =\Omega \mid \xi \mid ^2$. Inserting the latter
into (6.6) and integrating, we obtain, that the function $\alpha _{k-1}$ is
of the same form with $\Upsilon _{k-1}=\vartheta ^1+\mid \beta \mid ^2\Omega
_k$ and $\Omega _{k-1},\Gamma _{k-1}$ given by (6.3) and (6.5), and 
\[
d_{k-1}=d_k+\hbar \left( \Omega _k\sigma +\Gamma _k\left( \sigma +2{\rm Re}%
\kappa _k\bar{\upsilon}+\nu ^1\mid \kappa _k\mid ^2\right) \right) . 
\]
Summing $\Sigma _{i=k}^K\left( d_{i-1}-d_i\right) $ and taking into account
that $d_K=0$ and $\Gamma _K=0$, we obtain (6.4) also for $k-1$ $\vrule %
height.9exwidth.8exdepth-.1ex$

\begin{lemma}
Let us assume that starting from the instant $k+1$, the controls $u_k$ are
chosen to be linear $u_{k^1}=-\lambda _{k^1}z_{k^1}$ with the coefficients
(6.2), where $z_{k^1},k^1>k$ depend linearly on the results of the
subsequent indirect measurement $\eta _{k+1},\ldots ,\eta _{K-1}$ by virtue
of the formula (5.11) with the initial condition $z_k=z$. Let also the
indirect measurements be described by the Gaussian distributions (5.9) up to
the $k$. Then the operator $\rho _k\left( \zeta ^k,\zeta \right) $, defined
in (4.9), has the following normal form 
\[
\rho _k\left( \zeta ^k,\zeta \right) =\lambda _k\left( \zeta ^k\right) + 
\]
\begin{equation}
+:\left( \Upsilon _k\mid u+\lambda _k\hat{x}_k\mid ^2+\mid \phi -\kappa
_{k+1}\gamma \mid ^2\Gamma _{k+1}\mid z-\hat{x}_k\mid ^2\right) r_k^o\left(
y_k\mid \zeta ^k\right) :,
\end{equation}
where 
\[
\lambda _k\left( \zeta ^k\right) =:\left( \Omega _k\mid \hat{x}_k\mid
^2+\left( \hbar \left( \Omega _k+\Gamma _k\right) \Sigma _k+d_k\right) {\bf 1%
}\right) r_k^o\left( y_k\mid \zeta ^k\right) :, 
\]
the operator $\hat{x}_k$, defined in (5.16), is linear with respect to $y_k$%
, and $r_k^o\left( \eta \mid \zeta ^k\right) $ is the distribution (5.14)
with the parameters $\nu ^o=\nu +\mu ^o$, where $\mu ^o=\max \left(
0,\varepsilon \right) $.
\end{lemma}

{\noindent \bf Proof } %
%
Indeed, the operator $\rho _k\left( \zeta ^{k+1}\right) $ similar to the
density operator $\rho \in {\cal R}_k$ is defined by the distribution $%
r_k\left( \eta ,\zeta ^{k+1}\right) =$ 
\begin{equation}
\int \int \alpha _k\left( \xi ,\zeta ^{k+1}\right) q_k\left( \xi -\phi \xi
_{k-1}-\beta u_{k-1},\eta -\gamma \xi _{k-1}-\delta u_{k-1}\right)
p_{k-1}\left( \xi _{k-1}\mid \zeta ^k\right) d\xi d\xi _{k-1}.
\end{equation}
It is a symbol of the contrary order (see the Appendix), which is normal
when $\varepsilon <0$ and antinormal when $\varepsilon >0$. In the former
case, inserting the operator symbol (6.4) into (6.9) and integrating with
respect to the Gaussian type of the distribution $p_{k-1}\left( \xi
_{k-1}\mid \zeta ^k\right) $, we obtain (6.7), where $r_k^o\left( \eta \mid
\zeta ^k\right) $coincides with the distribution $r_k\left( \eta \mid \zeta
^k\right) $of the Gaussian type (5.13) with the parameter $v^1=\nu $. In the
latter case $\varepsilon >0$, the normal symbol of the operator $\rho
_k\left( \zeta ^{k+1}\right) $ is obtained from (6.9) by means of the
convolution of type (A.2) with the distribution (5.9) with $\mu =\varepsilon 
$, and in the result of the parameter $\nu $ increases for $\varepsilon $.
In this case $r_k^o\left( \eta |\zeta ^k\right) $ is also the normal symbol
of the conditional density operator $\rho _k\left( \zeta ^k\right) $ on $%
{\cal B}_k$ $\vrule height.9exwidth.8exdepth-.1ex$

\begin{theorem}
Let the quantum oscillator (5.1), (5.2) be described by the Gaussian initial
and transitional distributions of the Gaussian form (5.7), (5.8), and the
quality criterion (4.2) be defined by the quadratic final and transitional
operators $\alpha _{K}=\Omega x_{K}^{\ast }x_{K}$ and $c_{k}\left(
u_{k}\right) $ of form (6.1) respectively. Then the optimal strategy is
linear: $u_{k}=-\lambda _{k}z_{k}$, where $\lambda _{k}$ is defined by
(6.2), and $z_{k}$ are optimal linear estimates (5.11) based on the results $%
\left\{ \eta _{i}\right\} _{i\leq k}$ of the coherent measurements (5.5)
which are described by the distributions (5.9) with the minimal value of the
parameter $\mu =\mu ^{o}$.
\end{theorem}

{\noindent \bf Proof } %
%
We should verify the necessary and sufficient optimality conditions (4.10),
(4.11) for the operator (6.7) and the mentioned above measurement at each
instant $k$. As $\Upsilon _k,\Gamma _k\ge 0$, and the density operator $%
:r_k\left( y_k\mid \zeta ^k\right) :$ is non--negative definite, the
differences $\rho _k\left( \zeta ^{k+1}\right) -\lambda _k\left( \zeta
^k\right) $ are non-negative definite operators as well. It remains to
verify the equations (4.13) for the optimal strategy $\gamma _k^o\left(
\zeta ^k,\eta \right) =-\lambda _kz_k$ of the coherent measurements (5.5)
or, what is the same, of the measurements (5.15) with the Gaussian
distributions $n_k^o\left( z\right) $, corresponding to the case $\mu =\mu
^o $. Inserting $u=-\lambda _kz$ into (6.7) and taking into account (6.5),
we obtain 
\[
\rho _k\left( \zeta ^k,\eta ,\gamma _k^o\left( \zeta ^k,\eta \right) \right)
-\lambda _k\left( \zeta ^k\right) =\Gamma _k:\mid z-\hat{x}_k\mid
^2r_k\left( y_k\mid \zeta ^k\right) :. 
\]
Thus, equations (4.13) with $\varepsilon >0$ can be written in the form 
\begin{equation}
\left( z-\hat{x}_k\right) \#n_k^o\left( z-\hat{x}_k\right) \#=0,
\end{equation}
and the adjoint ones can be written for $\varepsilon <0$ also as 
\begin{equation}
\#n^o\left( z-\hat{x}_k\right) \#\left( z-\hat{x}_k\right) =0.
\end{equation}
The operators $\#n_k^o\left( z-\hat{x}_k\right) \#$ described by the
Gaussian distributions $n_k^o\left( z\right) $, which realize the lower
bound of the Heisenberg inequality, are well known as proportional to
coherent projectors [8]. The operators $\hat{x}_k$ when $\varepsilon >0$,
are proportional to the annihilation operators, and when $\varepsilon <0$,
they are proportional to the creation operators, for which the coherent
projectors are the right and the left eigen-projectors respectively. Hence,
the equations (6.9) is satisfied in the case $\varepsilon >0$, and the
equation (6.10) is satisfied if $\varepsilon <0$. Note that, in the
antinormal case when the coherent projectors are described by the Dirac
distributions $\delta \left( z\right) $ on ${\bf C}$, these equations are
written as the identities 
\[
\left( z-\hat{x}_k\right) \#\delta \left( z-\hat{x}_k\right) \#=\#\left( z-%
\hat{x}_k\right) \delta \left( z-\hat{x}_k\right) \#=0,\quad \varepsilon >0, 
\]
\[
\#\delta \left( z-\hat{x}_k\right) \#\left( z-\hat{x}_k\right) =\#\delta
\left( z-\hat{x}_k\right) \left( z-\hat{x}_k\right) \#=0,\quad \varepsilon
<0. 
\]

The minimal losses, corresponding to the optimal quantum strategy are
defined by the following formula 
\begin{equation}
\alpha ^o=\Omega _0\mid z\mid ^2+\hbar \left( \Omega _0\Sigma
+\sum_{k=1}^K\left( \Omega _k\sigma +\bar{\lambda}_{k-1}\right) \left( \phi 
\bar{\beta}\Omega _k-\vartheta \right) \Sigma _{k-1}\right) ,
\end{equation}
where $\lambda _k,\Omega _k$ are defined by (6.2), (6.3), and $\kappa
_k,\Sigma _k$ by (5.11), (5.12) with $\mu =\max \left( 0,\varepsilon \right) 
$ $\vrule height.9exwidth.8exdepth-.1ex$

Let us also obtain the solution to the corresponding time-continuous optimal
control problem for the quantum open system, described by the linear
stochastic differential equations (5.17), (5.18) and the quadratic integral
criterion 
\[
\Omega \left( :\mid x\left( \tau \right) \mid ^2:\right) +\int_0^\tau \left(
\omega :\mid x\left( t\right) \mid ^2:-2{\rm Re}\vartheta \bar{u}\left(
t\right) x\left( t\right) +\vartheta ^1\mid u\left( t\right) \mid ^2\right)
dt. 
\]
This criterion is obtained by setting $\omega \simeq \omega \Delta ,\quad
\vartheta \simeq \vartheta \Delta ,\quad \vartheta ^1\simeq \vartheta
^1\Delta $ in the conditions of the Theorem 6.1$,$ and passing to the limit
as $\Delta \longrightarrow 0$. So, the solution to the quantum optimal
control problem for the time continuous quantum open system (5.17), (5.18)
with quantum white noises $v,w$ is defined as the limit of the solution to
the discrete problem at $\Delta \longrightarrow 0$.

The optimal strategy, obtained in this limit, is obviously linear with
respect to the optimal estimate $z\left( t\right) $ of $x\left( t\right) $
as in the classical case [20]: $u\left( t\right) =-\lambda \left( t\right)
z\left( t\right) $, where $\lambda \left( t\right) =\left( \bar{\beta}\Omega
\left( t\right) -\vartheta \right) /\vartheta ^1,\quad \Omega \left( \tau
\right) =\Omega ,$ and $\Omega \left( t\right) $ satisfies the equation: 
\[
-d\Omega \left( t\right) /dt+\left( \alpha +\bar{\alpha}\right) \Omega
\left( t\right) =\omega -\mid \lambda \left( t\right) \mid ^2\vartheta ^1. 
\]
The optimal estimate $z\left( t\right) $ is obtained by coherent
measurements, corresponding to the case $\mu =\max \left( 0,\varepsilon
\right) $ in the time-continuous Kalman filter, and the minimal mean square
losses are defined by the integral 
\[
\alpha ^o=\Omega _0\mid z\mid ^2+\hbar \left( \Omega _0\Sigma +\int_0^\tau
\left( \Omega \left( t\right) \sigma +\bar{\lambda}\left( t\right) \right)
\left( \phi \bar{\beta}\Omega \left( t\right) -\vartheta \right) \Sigma
\left( t\right) dt\right) . 
\]
In particular, when $\beta =\gamma =\varepsilon =\alpha +\bar{\alpha}%
>0,\quad \nu =\upsilon =\sigma ,\quad \vartheta ^1=\vartheta +\theta ,\quad
\vartheta =\omega $, we obtain the solution to the optimal control problem
for the quantum open oscillator matched with the transmission line (2.3) of
the wave resistance $\gamma /2$ which was considered as the motivating
example in \S 2. In this case the equations (5.17), (5.18) are reduced to
(2.2), (2.3), where the generalized derivatives $v\left( t\right) =v\left(
dt\right) /dt,\quad y\left( t\right) =y\left( dt\right) /dt$ represent the
direct and reverse waves on the input of the open oscillator.

\appendix 

\section{APPENDIX}

\setcounter{equation}{0} Let ${\cal A}$, ${\cal B}$ be von--Neumann
algebras, i.e. selfadjoint weakly closed subalgebras of operators in a
complex Hilbert space ${\cal H}$ including the identity operator ${\bf 1}$,
and ${\cal P}$, ${\cal R}$ be predual spaces of ultra weakly continuous
functionals on ${\cal A}$ and ${\cal B}$, respectively. The elements $\pi
\in {\cal P}$ and $\rho \in {\cal R}$ are called states on ${\cal A}$ and $%
{\cal B}$ respectively if $<\pi ,a>\,\ge 0$, $<\rho ,b>\,\geq 0\quad \forall
a\geq 0,b\geq 0$ ($a,b\geq 0$ means the non-negative definiteness of the
operators $a\in {\cal A}$ and $b\in {\cal B}$), and if $<\pi ,{\bf 1>}=1$, $%
<\rho ,{\bf 1>}=1$. Linear operators transforming operators $b\in {\cal B}$
into operators $a\in {\cal A}$ are called superoperators, and the predual
linear maps ${\cal P}\rightarrow {\cal R}$ are called operations. The
typical example of a superoperator gives a representation $b\mapsto u^{*}bu$%
, where $u$ is a unitary operator. An operation $M:\pi \mapsto \pi M\in 
{\cal R}$ is called the (statistical) morphism if the dual superoperator $%
b\mapsto Mb\in {\cal A}$ is positive\footnote{%
For a physical realization of the statistical morphisms by conditional
expectations of the representations a stronger condition of complete
positivity $\left[ Mb_{ik}\right] _{i,k=1...n}\geq 0,\forall n$, where $%
\left[ b_{ik}\right] _{i,k=1...n}\geq 0$ is any non-negative definite
operator-matrix with the elements $b_{ik}\in {\cal B}$, should be imposed on
the morphisms.} $Mb\geq 0,\quad \forall b\geq 0$ and $M{\bf 1}={\bf 1}$ (it
is convenient to denote the morphisms and dual superoperators by the same
symbol with the right and the left action respectively: $<\pi M,b>=<\pi ,Mb>$%
.)

A ${\cal B}$--valued measure $b\left( d\zeta \right) $ on some Borel space $%
Z\ni \zeta $ is called $Z$--measurement, if $b\left( d\zeta \right) \geq 0$
for any Borel $d\zeta \subseteq Z$ and $\int b\left( d\zeta \right) ={\bf 1}$
in the same sense. If $M:{\cal P}\rightarrow {\cal R}$ is a morphism
describing a quantum channel, $\pi $--the state on its input and $b\left(
d\zeta \right) $ --the measurement on its output, then the probability
distribution on $Z$ is calculated by any of the formulas 
\begin{equation}
P\left( d\zeta \right) =<\pi M,b\left( d\zeta \right) >=<\pi ,Mb\left(
d\zeta \right) >.
\end{equation}

Let, for instance, the subalgebras ${\cal A}$ and ${\cal B}$ be generated by
the operators $x$ and $y$ respectively with the canonical commutation
relations 
\[
\left[ x,y\right] =0,\quad \left[ x,x^{*}\right] =\hbar {\bf 1},\quad \left[
y,x^{*}\right] =\gamma \hbar {\bf 1},\quad \left[ y,y^{*}\right]
=\varepsilon \hbar {\bf 1}, 
\]
where $\gamma \in {\bf C},\varepsilon \in {\bf R}$ and $\hbar >0$ is a
constant.

It may be assumed that $y=\gamma x+v$ holds, where $v$ is an operator in $%
{\cal H}$ commuting with $x$ and $x^{*}$, but not commuting with the adjoint
one: $\left[ v,v^{*}\right] =\left( \varepsilon -\mid \gamma \mid ^2\right)
\hbar {\bf 1}$, and the algebra generated by the pair $x,y$ can be
represented in the form of the tensor product ${\cal A}\otimes {\cal B}%
^{\circ }$, where ${\cal B}^{\circ }$ is the von--Neumann algebra generated
by the operator $v$.

We shall write the operators, generated by the operators $x$ and $v$ in the
form $\#\varphi \left( x,v\right) \#$, where $\varphi \left( \xi ,\eta
\right) $ are complex--valued functions of $\xi ,\eta \in {\bf C}$, called
symbols, and the notation $\#\cdot \#$ indicates such order of action for
the operators between them, that first act the operators $x,v$, and then
their conjugate. For instance, $\#\mid x\mid ^2\#=x^{*}x$. In a sufficiently
wide class of symbols any operator from ${\cal A}\otimes {\cal B}^{\circ }$
can be represented in such a form, and this representation is single-valued
and injective. In the case $y=\gamma x+v$ the operators $a\in {\cal A}$ are
described by the symbols $\varphi \left( \xi ,\eta \right) =\alpha \left(
\xi \right) $ and the operators $b\in {\cal B}$ by the symbols $\varphi
\left( \xi ,\eta \right) =\beta \left( \gamma \xi -\eta \right) $as in the
classical commutative case $\hbar =0$. The states in this quasi-classical
representation are described by distributions $q\left( \xi ,\eta \right) $,
generalizing the probability densities and representing the density
operators as the symbols of the contrary order, which are dual to the order
for the symbols $\varphi \left( \xi ,\eta \right) $. Due to $\hbar >0$, $%
x^{*}/\sqrt{\hbar }$ is the standard creation operator, and $x/\sqrt{\hbar }$
is the standard annihilation operator, so that the representation $%
a=\#\alpha \left( x\right) \#$ of operators $a\in {\cal A}$ is normal [19],
described by the holomorphic symbols $\alpha \left( \xi \right) $ with
respect to both $\xi ,\bar{\xi}$. The corresponding symbols $p\left( \xi
\right) $ of the states $\pi $ on ${\cal A}$ are described by the Glauber
distributions $p\left( \xi \right) $, which are defined as the linear
functionals 
\[
<\pi ,a>=\int p\left( \xi \right) \alpha \left( \xi \right) d\xi \quad
\left( d\xi =d{\rm Re}\xi d{\rm Im}\xi /\pi \hbar \right) , 
\]
describing the symbols of the density operator $\pi $, appropriate to the
antinormal order. The normal order is denoted by the parentheses $\ :\quad :$
, so we have $\#\alpha \left( x\right) \#=:\alpha \left( x\right) :$ when $%
\left[ x,x^{*}\right] \ge 0$. Note, that the antinormal symbol $p\left( \xi
\right) $ of the density operator $\pi $ and the normal symbol $p^o\left(
\xi \right) $ are connected by the convolution [21] 
\begin{equation}
p^o\left( \xi \right) =\int \exp \left\{ -\mid \xi -\xi ^1\mid ^2/\hbar
\right\} p\left( \xi ^1\right) d\xi ^1.
\end{equation}

The appropriate representation of the algebra ${\cal B}^{\circ },$ and hence 
${\cal A}\otimes {\cal B}^{\circ }$, is normal only if $\varepsilon >\mid
\gamma \mid ^2$, when $\left[ v,v^{*}\right] >0$. If $m\left( \eta \right) $
is a distribution which defines a state on ${\cal B}^{\circ }$ and there is
no statistical dependence, a state on ${\cal A}\otimes {\cal B}^{\circ }$ is
described by the product $p\left( \xi \right) m\left( \eta \right) $ and a
state on the sub-algebra ${\cal B}$ by the convolution 
\begin{equation}
r\left( \eta \right) =\int m\left( \eta -\gamma \xi \right) p\left( \xi
\right) d\xi .
\end{equation}
A superoperator ${\cal B}\rightarrow {\cal A}$, which is dual to a morphism
(A.3), is described by the symbol transformation 
\[
\alpha \left( \xi \right) =\int \beta \left( \eta \right) m\left( \eta
-\gamma \xi \right) d\eta . 
\]

For the normality of the appropriate representation $b=\#\beta \left(
y\right) \#$ of the operators $b\in {\cal B}$ with the distribution (A.3)
being Glauber, it is sufficient, that $\varepsilon >0$. When $\varepsilon <0$%
, the distribution $r\left( \eta \right) $ is the normal symbol of the
appropriate density operator $\rho =\#r\left( y\right) \#$.

Let us consider the complex measurements, described by the measurements of
the sum $\kappa y+w=z$, where $w$ is an operator in ${\cal H}$, which
commutes with $y$ and $y^{\ast }$, but does not commute with the adjoint one 
$w^{\ast }$: 
\[
\left[ w,w^{\ast }\right] =-\varepsilon \mid \kappa \mid ^{2}\hbar {\bf 1},
\]
so that $\left[ z,z^{\ast }\right] =0$ (it is assumed that the space ${\cal H%
}$ is chosen sufficiently wide, otherwise such an operator in ${\cal H}$ may
not exist.)

If $\eta \left( \zeta \right) $ is a distribution describing a state on the
algebra ${\cal B}^1$ generated by the operator $w$, then the probability
distribution of the results of such a measurement on the output of the
channel is described by the normed with respect to the Lebesgue measure $%
d\zeta $ density 
\[
s\left( \zeta \right) =\int \int n\left( \zeta -\kappa \eta \right) m\left(
\eta -\gamma \xi \right) p\left( \xi \right) d\eta d\xi . 
\]
In accordance with formula (A.1) such a measurement is described by the $%
{\cal B}$ --valued measure 
\begin{equation}
b\left( d\zeta \right) =\#n\left( \zeta -\kappa y\right) \#d\zeta ,
\end{equation}
and the distribution $n\left( \zeta \right) $ satisfies the condition 
\begin{equation}
\int \mid \zeta \mid ^2n\left( \zeta \right) d\zeta \geq \max \left\{
\varepsilon \mid \kappa \mid ^2\hbar ,0\right\}
\end{equation}
in accordance with the inequality $w^{*}w\geq \left\{ \left[ w^{*},w\right]
,0\right\} $. When $\varepsilon >0$ and representation (A.4) is normal,
inequality (A.5) prohibits, in particular, distributions of Dirac $\delta $
--form.

\newpage\ 

\begin{center}
REFERENCES
\end{center}

\smallskip\ 

\begin{enumerate}
\item  C.\thinspace W.\thinspace Helstrom. ``Detection theory and quantum
mechanics''. Inf. Contr., vol. 10, pp.\thinspace 254-291, Mar. 1967.

\item  H.\thinspace P.\thinspace Yuen, K.\thinspace S.\thinspace Kennedy and
M.\thinspace Lax. ``On optimal quantum receiver for digital signal
detection. Proc.\thinspace IEEE, vol. 58, pp.\thinspace 1770-1773, 1970.

\item  V.\thinspace P.\thinspace Belavkin and B.\thinspace A.\thinspace
Grishanin. ``Opitmal measurement of quantum observables''. Problems of
inform. trans., vol.\thinspace 8, pp.\thinspace 103-109, 1972 in Russian.

\item  A.\thinspace S.\thinspace Holevo. ``Statistical problems in quantum
physics'', in Proc.\thinspace Soviet--Japanese Symp.\thinspace on
Probability and Statistics, vol.\thinspace 1, pp.\thinspace 22-40, 1972.

\item  R.\thinspace L.\thinspace Stratonovich. ``The quantum generalization
of optimal statistical estimation and hypothesis testing''. Stochastics,
vol.\thinspace 1, pp.\thinspace 87-126, 1973.

\item  V.\thinspace P.\thinspace Belavkin. ``Optimal multiple quantum
statistical hypothesis testing''. Stochastics, vol.\thinspace 1, pp 315-345,
1975.

\item  C.\thinspace W.\thinspace Helstrom. ``Quantum detection and
Estimation Theory''. Academic Press, New York, San Francisco, London, 1976.

\item  V. P.Belavkin ``Nondemolition measurement and control in quantum
dynamical systems''. CISM Courses and Lectures, 294, pp. 311-329, Springer,
Vienna, 1987.

\item  V.P.Belavkin. ``Nondemolition measurement and nonlinear filtering of
quantum stochastic processes''. Lecture Notes in Control and Information
Sciences, 121, pp. 245-266, Springer Verlag, 1988.

\item  V.P.Belavkin. ''Quantum filtering of Markovian signals with quantum
white noises''. Radiotechnika and Electronika, 25: 1445, 1980. ( The English
translation is in: ''Quantum Communication and Measurement, Ed. by V.P.
Belavkin at al, pp. 381-391, Plenum Publisher, New York \& London, 1995.)

\item  V.P.Belavkin. ``Optimal measurement and control in quantum dynamical
systems''. Preprint no 411, Inst. of Phys., Copernicus University,
Toru\'{n}, February 1979.

\item  V.\thinspace P.\thinspace Belavkin. ``Quantum stochastic calculus and
quantum nonlinear filtering''. J. Multivariate Analysis, 42 (2), pp.171-201,
1992..

\item  V.\thinspace P.\thinspace Belavkin. ``Optimal quantum filtering of
Markov signals''. Problems of Control and Inform.\thinspace theory,
vol.\thinspace 5, 1978.

\item  P.\thinspace L.\thinspace Stratonovich. ``Conditional Markoff
processes and their applications to optimal control''. Moscow state
university, Moscow, 1966.

\item  R.\thinspace Bellman, Dynamic Programming, Princeton University
Press, Princeton, N.\thinspace Y. 1957.

\item  H.\thinspace A.\thinspace Haus. ``Steady-state quantum analysis of
linear systems''. Proc.\thinspace IEEE, vol.\thinspace 58, pp.\thinspace
1599-1611, 1970.

\item  M.\thinspace Lax. ``Quantum noise IV''. Quantum theory of noise
sources. Phys. Rev., vol.\thinspace 145, pp.\thinspace 110-129, 1965.

\item  Von Neumann J. ``Mathematical foundation of quantum mechanics''.
Princeton Univ.\thinspace Press, Princeton, N.\thinspace Y.\thinspace 1955.

\item  G.\thinspace Emch. ``Algebraic methods in statistical mechanics and
quantum field theory''. Wiley--Interscience, a division of John Wiley and
sons, inc.\thinspace New York, London, Sydney, Toronto, 1972.

\item  A.\thinspace Kossakowski. ``On quantum statistical mechanics of
non--Hamiltonian systems''. Rep.\thinspace Math.\thinspace Phys.,
vol.\thinspace 3 pp.\thinspace 247-274, 1972.

\item  J.\thinspace R.\thinspace Klauder and E.\thinspace C.\thinspace
D.\thinspace Sudarshan. ``Fundamentals of quantum optics''. W.\thinspace
\.{A} Benjamin, inc.\thinspace New York, Amsterdam, 1968.

\item  K.\thinspace J.\thinspace Astr\"{o}m. ``Introduction to stochastic
control theory. Academic Press, New York, 1970.
\end{enumerate}

\end{document}